\documentclass[sigconf]{acmart}

\AtBeginDocument{%
  }

\copyrightyear{2023}
\acmYear{2023}
\setcopyright{acmlicensed}
\acmConference[KDD '23]{Proceedings of the 29th ACM SIGKDD Conference on Knowledge Discovery and Data Mining}{August 6--10, 2023}{Long Beach, CA, USA}
\acmBooktitle{Proceedings of the 29th ACM SIGKDD Conference on Knowledge Discovery and Data Mining (KDD '23), August 6--10, 2023, Long Beach, CA, USA}
\acmPrice{15.00}
\acmDOI{10.1145/3580305.3599769}
\acmISBN{979-8-4007-0103-0/23/08}

\begin{document}

\title{AdaTT: Adaptive Task-to-Task Fusion Network for Multitask Learning in Recommendations}

\author{Danwei Li}
\authornote{Correspondence to lidli@meta.com.}
\email{lidli@meta.com}
\affiliation{%
  \institution{Meta AI}
  \streetaddress{1 Hacker Way}
  \city{Menlo Park}
  \state{California}
  \country{USA}
  \postcode{94025}
}

\author{Zhengyu Zhang}
\email{zhengyuzhang@meta.com}
\affiliation{%
  \institution{Meta Platforms, Inc.}
  \streetaddress{1 Hacker Way}
  \city{Menlo Park}
  \state{California}
  \country{USA}
  \postcode{94025}
}

\author{Siyang Yuan}
\email{syyuan@meta.com}
\affiliation{%
  \institution{Meta AI}
  \streetaddress{1 Hacker Way}
  \city{Menlo Park}
  \state{California}
  \country{USA}
  \postcode{94025}
}

\author{Mingze Gao}
\email{gaomingze@meta.com}
\affiliation{%
  \institution{Meta Platforms, Inc.}
  \streetaddress{1 Hacker Way}
  \city{Menlo Park}
  \state{California}
  \country{USA}
  \postcode{94025}
}

\author{Weilin Zhang}
\email{weilinzzz@meta.com}
\affiliation{%
  \institution{Meta AI}
  \streetaddress{1 Hacker Way}
  \city{Menlo Park}
  \state{California}
  \country{USA}
  \postcode{94025}
}

\author{Chaofei Yang}
\email{yangcf10@meta.com}
\affiliation{%
  \institution{Meta AI}
  \streetaddress{1 Hacker Way}
  \city{Menlo Park}
  \state{California}
  \country{USA}
  \postcode{94025}
}

\author{Xi Liu}
\email{xliu1@meta.com}
\affiliation{%
  \institution{Meta AI}
  \streetaddress{1 Hacker Way}
  \city{Menlo Park}
  \state{California}
  \country{USA}
  \postcode{94025}
}

\author{Jiyan Yang}
\email{chocjy@meta.com}
\affiliation{%
  \institution{Meta AI}
  \streetaddress{1 Hacker Way}
  \city{Menlo Park}
  \state{California}
  \country{USA}
  \postcode{94025}
}

\renewcommand{\shortauthors}{Li, et al.}

\begin{abstract}
Multi-task learning (MTL) aims to enhance the performance and efficiency of machine learning models by simultaneously training them on multiple tasks. However, MTL research faces two challenges: 1) effectively modeling the relationships between tasks to enable knowledge sharing, and 2) jointly learning task-specific and shared knowledge. In this paper, we present a novel model called Adaptive Task-to-Task Fusion Network (AdaTT)\footnote{The code is available at \url{https://github.com/facebookresearch/AdaTT}.} to address both challenges. AdaTT is a deep fusion network built with task-specific and optional shared fusion units at multiple levels. By leveraging a residual mechanism and a gating mechanism for task-to-task fusion, these units adaptively learn both shared knowledge and task-specific knowledge. To evaluate AdaTT's performance, we conduct experiments on a public benchmark and an industrial recommendation dataset using various task groups. Results demonstrate AdaTT significantly outperforms existing state-of-the-art baselines. Furthermore, our end-to-end experiments reveal that the model exhibits better performance compared to alternatives.
\end{abstract}



\maketitle

\section{Introduction}
Online recommendation systems aim to generate personalized high-quality recommendations to users. The effectiveness of these systems often depends on their ability to accurately learn user preferences, which typically requires optimizing multiple objectives simultaneously. For example, a short video recommendation system should consider both the likelihood of users watching a video and the likelihood of them liking it. Multi-task learning (MTL) is a typical solution for such use cases. By jointly training multiple tasks within a single framework, MTL offers several benefits. Firstly, it increases computational efficiency, which is important for large-scale online recommendation systems. Additionally, it enhances model performance through cross-task regularization and knowledge sharing.

However, MTL also poses unique challenges. One of the main challenges is modeling the relationships between tasks. Since each task may have varying degrees of correlation with others, merely modeling general commonalities shared by all tasks is insufficient. The complexity of this problem increases with the number of tasks. Effective task relationship modeling is the key to efficient task-adaptive knowledge sharing. For example, knowledge shared for the task "sharing a video" can be heavily weighted on similar tasks such as "liking a video", while also drawing on different aspects of knowledge from other tasks with abundant examples, such as "watching a video." On the other hand, it would minimize the shared learning with tasks that are highly irrelevant. Prior works \cite{liu2019end, caruana1997multitask} often resort to static shared representations. Other works like cross-stitch networks \cite{misra2016cross}, shown in Figure~\ref{fig:exp_models} (c), learn matrices to model relationships between multiple sub-networks. Yet, the weights remain fixed for all the examples and the sub-networks are only loosely task-specific.  Recent approaches like MMoE \cite{ma2018modeling} shown in Figure~\ref{fig:exp_models} (b) and PLE \cite{tang2020progressive} shown in Figure~\ref{fig:exp_models} (e) use specialized gating networks to dynamically combine shared sub-modules for flexible sharing, but the relationships between tasks modeled by these methods are obscure and indirect.

In addition to shared learning, task-specific learning is an integral part in multi-task learning. Striking the right balance between the two is important for resolving task conflicts and achieving cross-task regularization. On the one hand, MTL can suffer from negative transfer, where the optimization for one task negatively impacts the performance on another task, particularly when the tasks have conflicting objectives. In such scenarios, MTL models should adaptively emphasize task-specific learning. On the other hand, excessive task-specific learning and insufficient sharing can lead to overfitting, diminishing the benefits of cross-task regularization. The quantity and distribution of training data for each task also impacts the focus of learning: tasks with more data can rely more on their specific learning, while those with less data or highly skewed data can concentrate more on shared learning. Taking example differences into account can make the tradeoff between the two even more dynamic. Therefore, it is important to automatically learn to balance these two types of learning. Many soft parameter sharing models can accomplish this without the need for tedious manual tuning \cite{caruana1997multitask} or learning static structures for all examples with simplified assumptions \cite{vandenhende2019branched, meyerson2017beyond, sun2020adashare}. However, further research is needed to understand how to model the interaction between shared and task-specific learning to improve performance.

To jointly tackle these challenges, we propose a novel MTL model, Adaptive Task-to-Task Fusion Network (AdaTT). To improve shared learning and interpretability, we propose to introduce task-specific experts, shared experts and gating modules to explicitly model the task-to-task interaction at both the task-pair and all-task levels. For synergistic task-specific learning and shared learning, we distinguish and model them in separate fusion modules, with different experts and fusion strategies applied in each. The fused results are then combined by a residual mechanism \cite{he2016deep}. Furthermore, we employ multiple levels of fusion, each specialized for different functionality, to enhance learning performance.

To assess the performance of AdaTT, we conduct experiments on a real-world short video recommendation system. We vary the experiment groups to examine its adaptability to different task relationships. Additionally, we use a public benchmark to further demonstrate its generalizability. In all these experiments, AdaTT consistently outperforms the baseline models across different datasets and task groups.

To evaluate AdaTT's performance at scale, we conduct studies on its hyperparameters, specifically focusing on the number of fusion levels and experts. Additionally, we design an ablation study and a visualization analysis to gain insight into AdaTT's internal mechanism. The ablation study validates the effectiveness of the residual design, with separately modeled fusion modules, in achieving complementary task-specific and shared learning. The visualization of expert weights at deep and shallow fusion levels provides a deeper understanding of the distinct and meaningful sharing patterns learnt across different fusion levels, tasks, and task groups.

In summary, the contributions of this paper are as follows:

\begin{itemize}
    \item We propose a novel MTL model, Adaptive Task-to-Task Fusion Network (AdaTT), that simultaneously achieves adaptive task-to-task knowledge sharing and robust task-specific learning.
    \item With thorough experimentation on real-world benchmark data and a large-scale video recommendation system, we evaluate the effectiveness of AdaTT compared to various baselines.
    \item We demonstrate the interpretability of the model by conducting ablation studies on its individual fusion modules and investigating the operation of its fusion units for both shallow and deep knowledge.

\end{itemize}

\section{Related Work}
Multitask learning has broad applications in various fields, including computer vision \cite{zhang2014facial, misra2016cross, kokkinos2017ubernet, liu2019end}, natural language processing \cite{hashimoto2016joint, collobert2008unified}, speech recognition \cite{deng2013new}, robotics \cite{williams2008multi}, and recommendation systems \cite{ma2018modeling, tang2020progressive, zhao2019recommending, hadash2018rank}.
Many research studies have focused on developing innovative MTL architectures. These models can be divided into two categories: hard parameter sharing and soft-parameter sharing. \textbf{Hard parameter sharing} involves using a predefined model architecture in which certain layers are shared among all tasks, while other layers are specific to individual tasks. The shared-bottom model \cite{caruana1997multitask} is one of the most widely used models of the hard parameter approach. The model utilizes shared lower layers for representation learning and has task-specific layers on top of it. Multilinear Relationship Networks \cite{long2017learning} improve on this structure by imposing tensor normal priors on parameters of task specific layers. Another example is UberNet \cite{kokkinos2017ubernet}, which solves diverse low-, mid-, and high-level visual tasks jointly using an image pyramid approach. It processes each resolution in the pyramid with both task-specific layers and shared layers. Hard parameter sharing models typically have a compact structure, but require significant manual efforts to determine what to share and lack adaptability. Also over-sharing across irrelevant or conflicting tasks can lead to negative transfer, which can negatively impact model performance.

To better address these challenges, many \textbf{soft parameter sharing} MTL models have been proposed. Cross-stitch network \cite{misra2016cross} and sluice network \cite{ruder2019latent} use trainable parameters to linearly combine the outputs of each layer. However, the linear combination they apply is fixed and hence doesn’t fully reflect task relationship distinction on individual examples. Other works have been proposed to use attention or gating modules, conditioned on inputs, to dynamically combine or extract knowledge for each task. For example, MTAN \cite{liu2019end} employs attention modules to produce elementwise masks which extract task-specific knowledge from a shared representation. MMoE \cite{ma2018modeling} introduces a mixture of experts and employs gating networks to dynamically fuse them for each task. More recently, PLE \cite{tang2020progressive} is proposed to further enhance the flexibility of knowledge sharing. PLE explicitly introduces task-specific experts in conjunction with shared experts. Moreover, PLE proposes progressive separation routing with gating modules to fuse knowledge selectively and dynamically.  Among this line of works, PLE is the most relevant to ours. Differently, our work introduces two kinds of complementary fusion modules to separately model task-specific learning and shared learning. Also, in addition to explicitly introducing shared modules for learning commonalities across all tasks,  we leverage direct task-pair fusion, based on the input, to maximize the flexibility of knowledge sharing.

\textbf{Neural Architecture Search (NAS)} \cite{elsken2017simple, zoph2016neural, liu2018darts, liu2018progressive, real2019regularized} methods have been applied to Multi-Task Learning (MTL) to automatically learn model structures. Branched Multi-Task Networks \cite{vandenhende2019branched} generate a tree structure by clustering tasks based on affinity scores and assigning dissimilar tasks to the different branches. \cite{guo2020learning} utilizes Gumbel-Softmax sampling for the branching operation instead of pre-calculated affinity scores, enabling end-to-end training. The Soft Layer Ordering technique \cite{meyerson2017beyond} identifies the limitations of traditional fixed-order sharing approaches in MTL models and proposes learning task-specific scaling parameters to enable a flexible ordering of shared layers for each task. AdaShare \cite{sun2020adashare} learns a task-specific policy to select which layers to execute for every particular task. Sub-Network Routing (SNR) \cite{ma2019snr} splits shared layers into sub-networks and learns their connections with latent variables. NAS methods eliminate  a significant amount of manual work and improve the flexibility of sharing patterns in MTL models. However, as an exhaustive search of all possible model configurations is combinatorially complex, these methods often rely on simplified assumptions such as branching \cite{vandenhende2019branched, guo2020learning}, routing \cite{ma2019snr}, layer ordering \cite{meyerson2017beyond}, layer selecting \cite{sun2020adashare}, etc. to limit the search space. Additionally, the generated structures don't adjust to individual examples.

In addition to works focusing on MTL architecture design, another line of works aims to improve MTL optimization.  Uncertainty-based weighting \cite{kendall2018multi} learns each task’s weight based on task uncertainty. GradNorm \cite{chen2018gradnorm} controls different tasks’ gradient magnitudes to balance their training rates.  GradDrop \cite{chen2020just} probabilistically selects a sign and removes gradients of the opposite sign. Gradient surgery (PCGrad) \cite{yu2020gradient} projects conflicting task gradients to each other's normal plan. RotoGrad \cite{javaloy2021rotograd} manipulates both the magnitude and direction of task gradients to alleviate conflicts. \cite{sener2018multi} treats multitask learning as a multi-objective optimization problem with the goal of finding a Pareto optimal solution. \cite{wang2022can} introduces self-auxiliary losses with under-parameterized small towers to balance Pareto efficiency and cross-task generalization.  While these approaches can bring improvements, relying solely on them without a strong model architecture may limit the upper bound of model performance.

\begin{figure*}[ht]
\begin{center}
	\centering
	\footnotesize
	\includegraphics[scale=0.15]{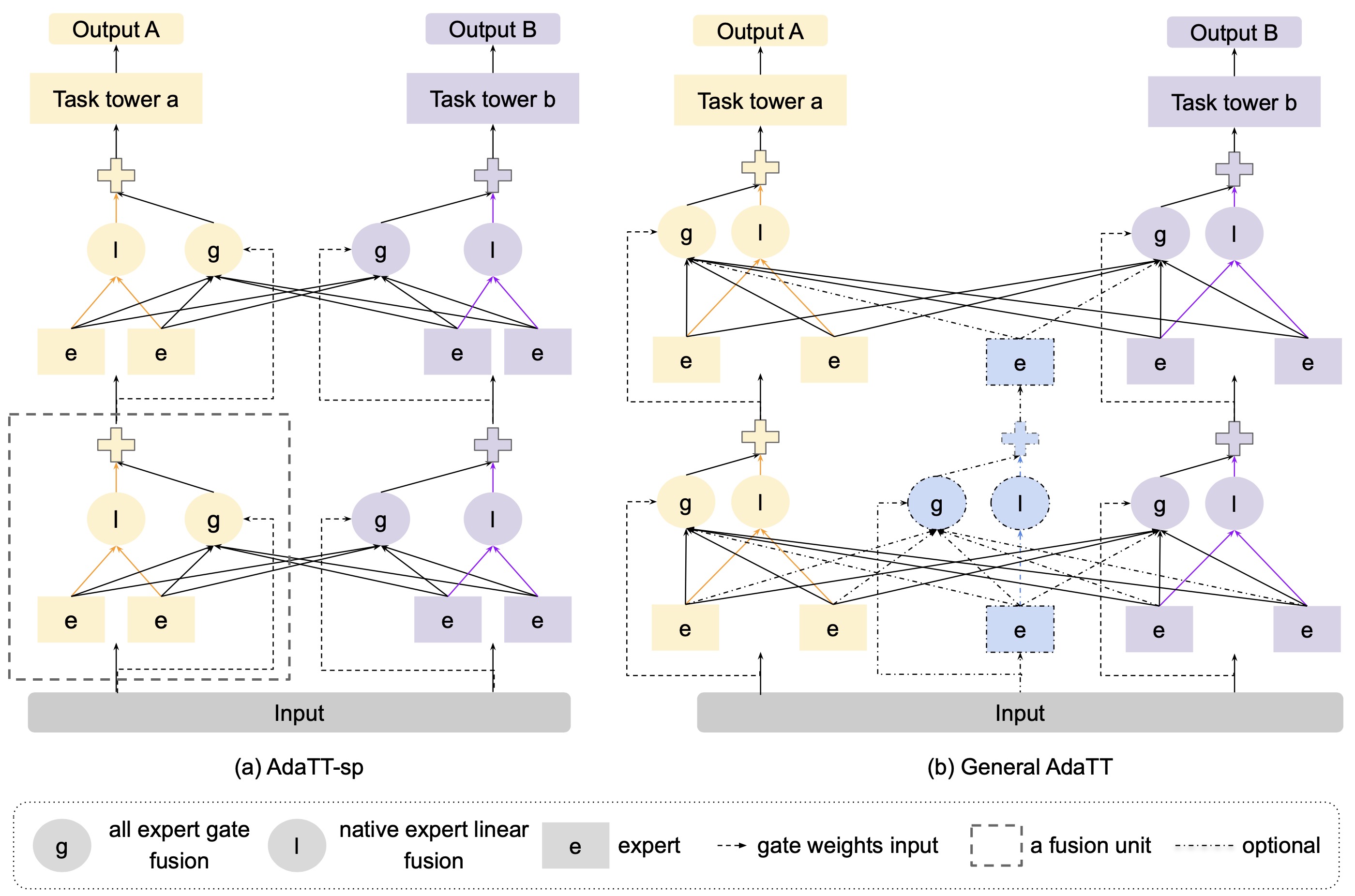}
\end{center}
\caption{AdaTT-sp and general AdaTT with 2 fusion levels. Task A and B's specific and shared modules are differentiated by color: yellow for A, purple for B, and blue for shared. For the purpose of illustration, we use 2 experts for each task-specific unit. In general AdaTT, we add a shared fusion unit which has a single expert as an example. Note that the shared modules in general AdaTT are not necessary and are therefore depicted using dotted lines. When there are no shared modules present, general AdaTT falls back to AdaTT-sp.}
\label{fig: AdaTT}
\end{figure*}

\section{Model architecture}
To learn adaptive shared representations and enhance task-specific learning jointly, we propose a new model, Adaptive Task-to-Task Fusion Network (AdaTT). AdaTT leverages gating and residual mechanisms to adaptively fuse experts in multiple fusion levels.
Consider a multi-task learning scenario with two prediction tasks. We illustrate the architecture of AdaTT in Figure~\ref{fig: AdaTT} using two fusion levels. AdaTT consists of a multi-level fusion network and task towers. The fusion networks are constructed with task-specific and optional shared fusion units, while the task towers are built on top of the fusion network and connected to the task-specific units in the final fusion level. Our framework is generic, supporting flexible choices of expert modules, task tower networks, gating modules, and a configurable number of experts and fusion levels. In the following sections, we will first introduce a special case of AdaTT, named AdaTT-sp, that uses only task-specific fusion units (as depicted in Figure~\ref{fig: AdaTT} (a)). Then, we will describe the general AdaTT design, as shown in Figure~\ref{fig: AdaTT} (b).

\subsection{AdaTT-sp}
The detailed design of AdaTT-sp is presented as follows.
Given input $x$ for $T$ tasks, the prediction for task $t$ ($t = 1, 2, \ldots, T$) is formulated as:
\begin{equation}
  y_t = h_t(f_t^L(x)),
\end{equation}
where $L$ is the number of fusion levels, $h_t$ represents task $t$'s task tower,
and $f_t^L$ denotes the function to produce the output of task $t$'s fusion unit at the $L$-th fusion level. Here, $f_t^L(x)$ is computed by applying fusion layer(s) from bottom to top using Equations~\ref{eq:L0} and ~\ref{eq:fusion}:

\begin{equation}
\label{eq:L0}
f_1^0(x) = f_2^0(x) = \dots = f_T^0(x) = x
\end{equation}
\begin{equation}
\label{eq:fusion}
    f_t^l(x) = FU_t^l (f_1^{l-1}(x), f_2^{l-1}(x), \dots, f_T^{l-1}(x)),\quad l=1\ldots L
\end{equation}
Here, FU represents a fusion unit.

\subsubsection{Fusion Unit}
Below we detail the construction of $FU_t^l$ introduced in Equation~\ref{eq:fusion}.
For task $t$, after receiving all the outputs from the previous fusion level, we first construct $m_t$ native experts for this task, denoted by $E_{t, i}^l$, using a function $e_{t, i}^l$ and input $f_t^{l-1}(x)$. That is,
\begin{equation}
E_{t,i}^l = e_{t, i}^l(f_t^{l - 1}(x)),
\end{equation}
where $i = 1, 2, \ldots, m_t$ and $E_{t,i}^l \in \mathbb{R}^{1 \times d^l}$. Each expert network at level $l$ produces a vector of length $d^l$. For easier notation, at level $l$, we use $E_t^l$ and $E^l$ to denote the vertical concatenation of the experts belonging to task $t$ and all experts across tasks, respectively. Concretely,
$E_t^l$ and $E^l$ are represented as:
\begin{equation}
E_t^l = [E_{t, 1}^l,E_{t, 2}^l, \ldots, E_{t, m_t}^l],
\end{equation}
\begin{equation}
E^l = [E_1^l, E_2^l, \ldots, E_T^l],
\end{equation}
where $E_t^l \in \mathbb{R}^{m_t \times d^l}$ and $E^l \in \mathbb{R}^{(m_1+m_2+\ldots+m_T) \times d^l}$.
In the above equations, $[,]$ represents operation of vertically stacking vectors or sub-matrices into a larger matrix.

As a task can have different degrees of correlation to other tasks, $FU_t^l$ directly models task-to-task knowledge fusion with a gating
module $AllExpertGF_t^l$ which combines all tasks' experts $E^l$. In addition, we leverage a lightweight linear combination $NativeExpertLF_t^l$ to fuse task $t$'s native experts, $E_t^l$. Conceptually, the gating module models shared learning and the linear combination of native experts models task-specific learning. Concretely, the output of task $t$'s specific unit at layer $l$ is formulated as:
\begin{equation}
\label{eq:residual}
f_t^l(x)=AllExpertGF_t^l(E^l, G_t^l)+NativeExpertLF_t^l(E_t^l),
\end{equation}
where gate weights $G_t^l$ is used to combine $E^l$. Relative to the representations in Equation ~\ref{eq:fusion}, $E^l$ depends on all $f_1^{l-1}(x), f_2^{l-1}(x), \ldots , f_T^{l-1}(x)$, while $G_t^l$ and $E_t^l$ solely depend on $f_t^{l-1}(x)$.\\
Specifically, in Equation~\ref{eq:residual}, the experts are fused as follows:
\begin{equation}
\label{eq:linearFusion}
NativeExpertLF_t^l(E_t^l)={v_t^l}^\intercal E_t^l,
\end{equation}

\begin{equation}
\label{eq:gateFusion}
AllExpertGF_t^l(E^l, G_t^l)={G_t^l}^\intercal E^l,
\end{equation}
where $E^l$ is multiplied by gates $G_t^l \in \mathbb{R}^{(m_1+m_2+\ldots+m_T) \times 1}$ generated by a function $g_t^l$ in $AllExpertGF$ while $E_t^l$ is combined simply by a learnt vector $v_t^l \in \mathbb{R}^{m_t \times 1}$ in $NativeExpertLF$.
When $m_1 = m_2 = \ldots = m_T = 1$, i.e., all fusion units only have one expert, $NativeExpertLF_t^l(E_t^l)$ falls back to $E_t^l$, assigning a unit weight to the native expert for simplicity. There are many design options for $g_t^l$. A common one is to use a single-layer MLP activated by softmax:
\begin{equation}
g_t^l(f_t^{l-1}(x)) = softmax(W_t^l{f_t^{l-1}(x)}^\intercal).
\end{equation}
Here, $W_t^l \in \mathbb{R}^{(m_1+m_2+\ldots+m_T) \times d^{l-1}}$ is a learnt matrix.

\subsubsection{Simplification}
For implementation efficiency, given Equation~\ref{eq:linearFusion} and Equation~\ref{eq:gateFusion}, we can actually pad ${v_t^l}^\intercal$ with zeros to match the size of ${G_t^l}^\intercal$, add the weights, and perform a single multiplication to combine all experts. Thus, Equation~\ref{eq:residual} can be simplified as:
\begin{equation}
f_t^l(x)=(pad({v_t^l}^\intercal)+{G_t^l}^\intercal)E^l.
\end{equation}
As we can see, the inclusion of the linear fusion module leads to a minimal increase in computation.

\subsection{General AdaTT}
In its general form, as shown in Figure~\ref{fig: AdaTT} (b), AdaTT employs optional shared fusion units. Conceptually, the fusion between task-specific module pairs models fine-grained sharing, while the fusion of task-specific and shared modules transfers broad knowledge that applies to all tasks. This leads to efficient and flexible task-to-task knowledge sharing.
The computation of general AdaTT is similar to AdaTT-sp, except for the final fusion level, where the shared fusion units don't perform any fusion operations and only produce expert outputs for task-specific fusion units to process.

In summary, AdaTT explicitly learns task specific knowledge and adaptively fuses it with shared knowledge. The fusion is task-adaptive as: 1. The gating modules learn the residual with respect to the tasks' native experts. 2. Each task-specific unit fuses experts using a specialized gating module conditioned on input (which is unique starting from the second fusion level). By allowing each task to directly and flexibly learn shared knowledge from other tasks, AdaTT offers greater flexibility compared to PLE which only relies on shared expert(s) as the media. Additionally, AdaTT can opt to use task-specific experts only. Unlike PLE, which processes all selected experts in a single gating module, AdaTT separately fuses native experts in a different linear fusion module within each fusion unit. This design enhances the robustness of task-specific learning after each level of fusion. Despite its simplicity, our experiments reveal that it outperforms PLE, which applies selection to experts in different fusion units and uses different routing paths to differentiate these experts.
\section{Experiments}
In this section, we present comprehensive experimental results to highlight the effectiveness of our proposed AdaTT model and provide a better understanding of it.

This section is divided into four parts. We first briefly describe baseline models in Section~\ref{label:baselines}. Secondly, we evaluate the effectiveness of AdaTT against state-of-the-art multi-task learning models through experiments on real-world industrial and public datasets. For the industrial dataset, we use three different groups of prediction tasks to examine the performance of these multi-task learning models in various scenarios. Results are shared in Section~\ref{label:reels} and Section~\ref{label:census}. Next, we present individual component studies in Section~\ref{label:ablation_module} and Section~\ref{label:visualization}. We ablate the $NativeExpertLF$ module to validate the importance of AdaTT’s residual design which incorporates separate modules to fuse different experts. We also visualize the expert weights learned in each task-specific unit to demonstrate how AdaTT learns proper interactions between tasks, which is essential for effective knowledge sharing. Finally, in Section~\ref{label:hyperparameters}, we conduct studies on AdaTT’s hyperparameters to understand the relationship between the number of fusion levels and experts, and the performance of AdaTT.

\subsection{Baseline Models}\label{label:baselines}
We employ Shared-bottom, MMoE, Multi-level MMoE (an extension of the original single-level MMoE), PLE, and Cross-stitch Networks as our baselines. Of these models, MMoE, PLE, and Cross-stitch Networks all utilize the soft-parameter sharing technique.

\begin{itemize}
    \item \textbf{MMoE} \cite{ma2018modeling}: This model learns a specialized gating module for each task to fuse multiple shared experts. Given $n$ expert modules $e_1, e_2, ..., e_n$, task $t$'s task tower module $h_t$ and gating module $g_t$, the prediction of task $t$ is computed as:
    \begin{equation}
    y_t = h_t(f_t(x)),
    \end{equation}
    where \begin{equation}
    f_t(x) = g_t(x)[e_1(x), e_2(x), ..., e_n(x)].
    \end{equation}
    Here, [,] represents vertically stacking vectors into a matrix.

    \item \textbf{Multi-level MMoE (ML-MMoE)}: This model extends the original single-level MMoE by incorporating multiple levels of fusion. In ML-MMoE, higher level experts use the lower level experts that are fused by different gating modules as inputs. Similar to the original MMoE, all gating modules are conditioned on the same raw input.
    \item \textbf{Cross-Stitch} \cite{misra2016cross}: This model introduces cross-stitch units to linearly combine different tasks’ hidden layers with learnt weights.
    \item \textbf{PLE} \cite{tang2020progressive}: This model explicitly introduces both task specific and shared experts with a progressive separation routing strategy. Gating modules are used to fuse selected experts in both task-specific and shared units. In PLE, shared units can fuse all experts at the same level, while task-specific units only fuse their native experts and shared experts. This model is the closest to AdaTT.
\end{itemize}

\begin{figure*}[ht]
\begin{center}
	\centering
	\footnotesize
	\includegraphics[scale=0.1625]{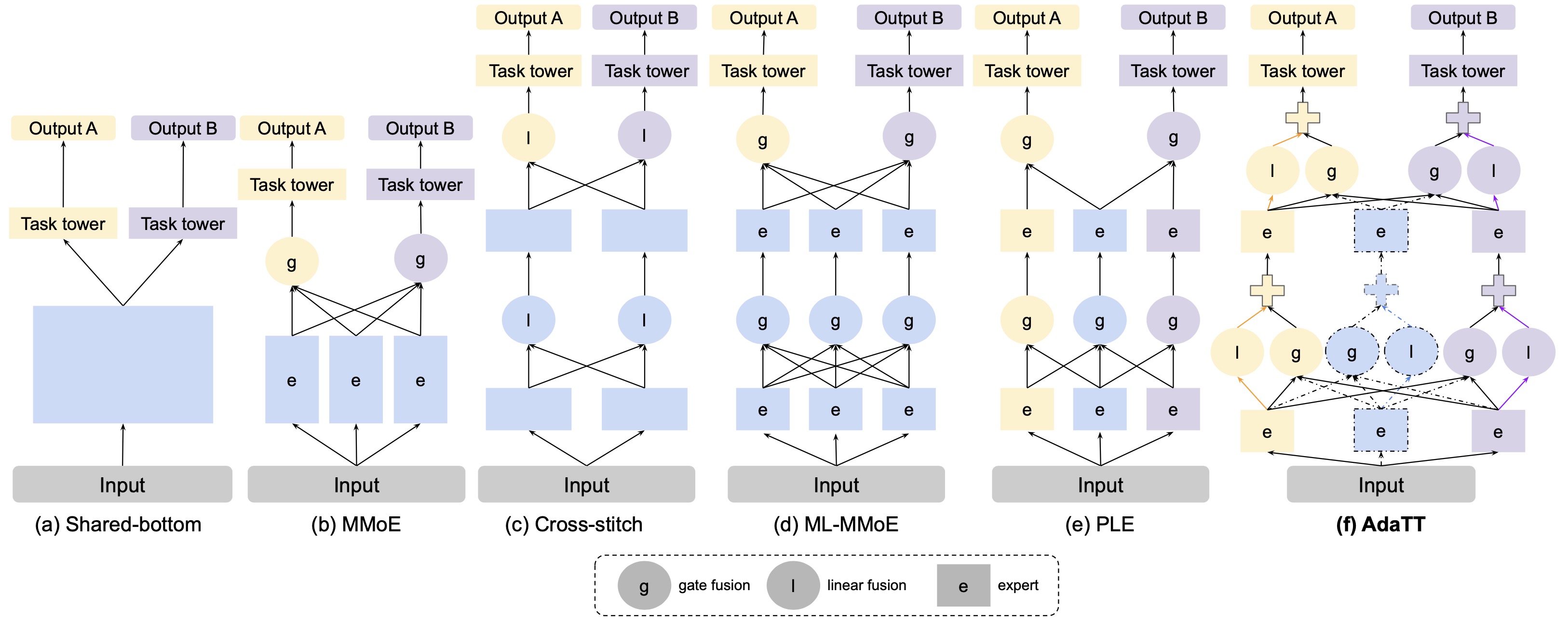}
\end{center}
\caption{MTL models used in our experiments. In multi-level MTL models, two fusion levels are used to illustrate their design. Modules are represented using different colors: shared modules are in blue, task A's specific modules are in yellow, and task B's specific modules are in purple.}  
\label{fig:exp_models}
\end{figure*}

All the models discussed above are shown in Figure~\ref{fig:exp_models} for comparison.

\subsection{Evaluation on Large-scale Short Video Recommendation}
\label{label:reels}
In this section, we present experimental results on a short video recommendation system. The system displays a list of recommended videos that are ranked based on scores from various tasks. These tasks can broadly be classified into two categories: engagement tasks, which take into account users' explicit feedback, such as commenting on a video, and consumption tasks, which reflect users' implicit feedback, such as video views.

\subsubsection{Task groups}
We create three groups of tasks to thoroughly assess these models' performance in different task relations.
\begin{itemize}
\item The first group includes one engagement task and one consumption task, which are expected to have a relatively low correlation.
\item Group two is made up of two consumption tasks that are more correlated. The first task is the same as the consumption task in group one. The second task is chosen to have a comparable positive event rate to the engagement task in group one. Both group 1 and group 2 are composed of binary classification tasks only.
\item In the third group, we increase the number of tasks to five and select highly diverse tasks. Out of these, three are consumption tasks, and two are engagement tasks. One of the consumption tasks is a regression task, and the remaining four tasks are binary classification tasks. In terms of user sentiments, we include one task reflecting users' dislike and four tasks for positive events. One of the engagement tasks with extremely sparse positive events is used as an auxiliary task.
\end{itemize}
When reporting results for all task groups, we present the regression task first (if present), followed by the classification tasks in the order of decreasing rate of positive examples.

\subsubsection{Experiment setup}
We collect a dataset of approximately 70 billion examples for training the models and test their performance on a test set of approximately 10 billion examples.
In feature processing, we convert all sparse features to dense embeddings and concatenate them with dense features. All tasks use the same input. All models are trained and tested using the same framework, with the same optimization settings such as the optimizer, learning rate, and batch size. For training, we use Cross Entropy loss for binary classification tasks and MSE loss for the regression task. The losses for all tasks are summed and optimized with equal weights. For testing, we use Normalized Entropy (NE) \cite{he2014practical} for binary classification tasks and MSE for the regression task.

\subsubsection{Model Hyperparameters}
\label{label:exp_hyperparameters}
In our experiments, all models have 3 hidden layers activated by ReLU. For each group of experiments, we conduct two comparisons.

First, we compare MMoE, PLE, and AdaTT to the shared-bottom model. For fair comparison, PLE, ML-MMoE, and AdaTT all have 2 levels of fusion. We use single-layer MLP experts with hidden dimensions of 256 and 128 for these 2 levels of fusion respectively. MMoE is constructed with 2-layer MLP experts with hidden dimensions of [256, 128]. We also set a limit on the total number of experts per level of fusion. All gating modules in these models use a 1-layer MLP with softmax activation. It's worth noting that the computation required for gating modules is significantly lighter compared to expert modules. Although both types of modules share the same input dimension, gating modules have a total output dimension that is nearly two orders of magnitude smaller. The task towers of all models have a single hidden layer of 64 units. With this setup, all models have comparable computation as the task towers and expert modules dominate the computation. In our experiments, we tune the number of task-specific and shared experts for PLE and AdaTT, while for MMoE, we tune the total number of experts.

In a separate experiment, we compare the performance of AdaTT and the cross-stitch model with the shared-bottom model. AdaTT uses similar hyperparameters as in the previous experiment but with 1 expert per task and no shared expert for comparability with the cross-stitch model. The cross-stitch model has 2 cross-stitch units and the same hidden layers as AdaTT.

\begin{table*}
  \caption{Performance on the consumption task in consumption + engagement task group}
  \label{tab:consum_eng1}
  \begin{tabular}{ccccc}
    \toprule
    Models&NE diff @10B examples&NE diff @30B examples&NE diff @70B examples& Test NE diff\\
    \midrule
    Shared-bottom & - & - & - & -\\
    MMoE & -0.334\% & -0.421\% & -0.498\% &-0.481\%\\
    ML-MMoE & -0.307\% & -0.400\% & -0.480\% &-0.463\%\\
    PLE & -0.162\% & -0.385\% & -0.482\% & -0.448\%\\
    AdaTT & \textbf{-0.391\%} & \textbf{-0.464\%} & \textbf{-0.526\%} & \textbf{-0.508\%}\\
    \midrule
    Cross-stitch & -0.024\% &-0.133\% &-0.166\% & -0.140\% \\
    AdaTT-sp (single task expert) & \textbf{-0.259\%} & \textbf{-0.261\%} & \textbf{-0.277\%} &\textbf{-0.231\%}\\
  \bottomrule
\end{tabular}
\end{table*}

\begin{table*}
  \caption{Performance on the engagement task in consumption + engagement task group}
  \label{tab:consum_eng2}
  \begin{tabular}{ccccc}
    \toprule
    Models&NE diff @10B examples&NE diff @30B examples&NE diff @70B examples& Test NE diff\\
    \midrule
    Shared-bottom & - & - & - & -\\
    MMoE & -0.370\% & -0.436\% &-0.542\% &-0.532\% \\
    ML-MMoE & -0.260\% & -0.386\% & -0.496\% &-0.494\%\\
    PLE & -0.360\% & -0.627\% & -0.698\% & -0.691\%\\
    AdaTT & \textbf{-0.677\%} & \textbf{-0.795\%} & \textbf{-0.845\%} & \textbf{-0.863\%}\\
    \midrule
    Cross-stitch & -0.046\% &-0.197\% &-0.232\% & -0.225\% \\
    AdaTT-sp (single task expert) & \textbf{-0.393\%} & \textbf{-0.367\%} & \textbf{-0.397\%} &\textbf{-0.362\%}\\
  \bottomrule
\end{tabular}
\end{table*}

\subsubsection{Experiment on the task group of engagement and consumption tasks}
For this group of tasks, we present the results of the NE difference for each model relative to the shared-bottom model, after training on 10 billion, 30 billion, and 70 billion examples. We also provide the test results. Table~\ref{tab:consum_eng1} and ~\ref{tab:consum_eng2} show the results for the consumption and engagement tasks respectively. The results indicate that AdaTT outperforms all other models in both tasks, achieving not only faster convergence but also higher quality. After training on 10 billion examples, the two AdaTT models already demonstrate a significant NE improvement for both tasks. As for the baseline models, PLE takes much longer to converge on the consumption task. The cross-stitch model, on the other hand, is outperformed by AdaTT by a large margin, demonstrating the vital importance of adaptive fusion in task relationship modeling.
Notably, PLE and AdaTT show greater improvement on the engagement task, which has fewer positive events, compared to the consumption task. However, this trend is not evident in MMoE and ML-MMoE, which highlights the importance of task-specific learning. Interestingly, despite having more flexibility through additional fusion operations, ML-MMoE performs worse than MMoE in both tasks, indicating its inferior performance in expert fusion. This is likely due to the lack of distinction and imposed prior knowledge in ML-MMoE’s design. The shared experts are highly symmetric, which are all used by each gating module, and there are no explicitly modeled task specific experts. Additionally, all the gating modules receive the same raw input. The increase in fusion levels results in more routes, making it more challenging for ML-MMoE to learn different weight combinations for predicting each particular task.

\subsubsection{Experiment on the task group of two consumption tasks}
As MTL models' performance can be sensitive to task correlations, we design an experiment group to evaluate their performance on two related consumption tasks in contrast to task group 1, where the correlation between tasks is lower. The results, as shown in Table~\ref{tab:taskGroup2}, reveal that all models in this group have more similar improvements on both tasks compared to the baseline. This is unsurprising, as when tasks are more closely related, negative transfer is less severe, and both tasks benefit from a higher level of shared knowledge. Even MTL models with simpler sharing mechanisms can achieve good performance, resulting in less prominent differences in NE. However, AdaTT still shows the best results among all the MTL models.

\begin{table}
  \caption{Performance on the consumption task group}
  \label{tab:taskGroup2}
  \begin{tabular}{ccc}
    \toprule
    Models & task 1 NE diff & task 2 NE diff\\
    \midrule
    Shared-bottom & - & -\\
    MMoE & -0.343\% & -0.372\% \\
    ML-MMoE & -0.415\% & -0.372\% \\
    PLE & -0.446\% & -0.368\% \\
    AdaTT & \textbf{-0.487}\% & \textbf{-0.443}\% \\
    \toprule
    Cross-stitch & -0.170\% & -0.136\% \\
    AdaTT-sp (single task expert) & \textbf{-0.233\%} & \textbf{-0.194\%} \\
  \bottomrule
\end{tabular}
\end{table}

\subsubsection{Experiment on five diverse tasks}
\label{label:FiveTasks}

In this group of tasks, we evaluate the models' abilities to handle complex cross-task relationships by utilizing 5 highly diverse tasks. We tune the models for the 4 main tasks and present the results in Table 4. We do not include the auxiliary task with sparse positive events due to its high level of noise and inconsistent performance. The results demonstrate that AdaTT outperforms all comparison models by a significant margin in all main tasks, indicating its superiority in handling complex task relationships.

\begin{table*}
  \caption{Model performance on the group of 5 tasks}
  \label{tab:freq}
  \begin{tabular}{ccccc}
    \toprule
    Models&Consumption task 1&Consumption task 2 &Consumption task 3 &Engagement task 1\\
    ~&MSE diff & NE diff & NE diff &NE diff\\
    \midrule
    Shared-bottom & - & - & - & -\\
    MMoE & -0.770\% & -0.632\% &-0.708\% &-1.182\% \\
    ML-MMoE & -0.697\% & -0.608\% & -0.685\% &-1.013\%\\
    PLE & -0.697\% & -0.599\% & -0.698\% & -1.221\%\\
    AdaTT & \textbf{-0.873\%} & \textbf{-0.738\%} & \textbf{-0.815\%} & \textbf{-1.346\%}\\
    \midrule
    Cross-stitch & -0.520\% &-0.454\% &-0.486\% & -0.818\% \\
    AdaTT-sp (single task expert) & \textbf{-0.613\%} & \textbf{-0.543\%} & \textbf{-0.589\%} &\textbf{-0.930\%}\\
  \bottomrule
\end{tabular}
\end{table*}

\subsection{Evaluation on a Public Dataset}\label{label:census}
\subsubsection{Dataset description} We use the Census Income dataset \cite{Dua:2019} extracted from the 1994 and 1995 current population surveys. The dataset has 40 features and 299,285 instances, including 199,523 training examples and 99,762 test examples. We randomly split the test examples into validation and test sets in equal ratio. The tasks are: 1) predicting whether the income exceeds 50K; 2) predicting whether the marital status is never married; 3) predicting whether the education is at least college.

\begin{table*}
  \caption{Performance on 3 tasks of UCI Census income dataset. We compare PLE, ML-MMoE and AdaTT using 2-level fusion. Expert and task tower networks are single-layer MLPs and their hidden dimensions are listed. The AdaTT-sp setup, which solely utilizes task-specific experts, enables AdaTT to achieve its optimal results.}
  \label{tab:census}
  \begin{tabular}{cccccccc}
    \toprule
    Model& total experts&expert hidden dimensions&task tower hidden dimension& task 1 AUC&task 2 AUC&task 3 AUC\\
    \midrule
    ML-MMoE & 6& 128, 64& 32&0.8729 & 0.9178 & 0.9731\\
    PLE & 6& 128, 64& 32&0.8683 & 0.9164 & 0.9697\\
    AdaTT &6& 128, 64&32&\textbf{0.8766} & \textbf{0.9202} & \textbf{0.9783}\\
    \midrule
    ML-MMoE &9& 96, 48& 32&0.8688 & 0.9139 & 0.9730\\
    PLE &9& 96, 48& 32&0.8645 & 0.9134 & 0.9680\\
    AdaTT &9& 96, 48&32& \textbf{0.8744} & \textbf{0.9174} & \textbf{0.9786}\\
  \bottomrule
\end{tabular}
\end{table*}

\subsubsection{Model hyperparameters}
This experiment employs a framework, adapted from \cite{AokiBIOKDD21}, to train and test ML-MMoE, PLE and AdaTT. Model structures are similar to those in Section~\ref{label:exp_hyperparameters}, but the hidden dimensions and number of experts are changed. The experiments are conducted in two groups with 6 and 9 experts per fusion level, respectively. $m_s$, the number of shared experts for PLE and AdaTT, is tuned. The number of task-specific experts is calculated as ${6-m_s}$ and ${9-m_s}$. To ensure fairness, all other hyperparameters are kept equal across the models. After tuning $m_s$, each model is trained 100 times with different initializations, and the mean AUC in the test set is reported.

\subsubsection{results}
Results are presented in Table~\ref{tab:census}. AdaTT outperforms the baseline models in all the tasks.

\subsection{Ablation study of the $NativeExpertLF$ module}\label{label:ablation_module}
In this section, we examine the effect of the residual mechanism with the $NativeExpertLF$ module in fusion units. We ablate the $NativeExpertLF$ module and only utilize the $AllExpertGF$ module to combine the outputs of all experts at each fusion unit.  We adopt a model structure similar to that of Section~\ref{label:exp_hyperparameters}, and use a fixed number of three experts per task and no shared expert.  Both models are trained on 70 billion examples and tested on 10 billion examples. The results are displayed in Table~\ref{tab:ResAblation}.

While the $AllExpertGF$ module can theoretically learn flexible expert combinations, our experiment nevertheless shows it is important to separately combine native experts and add the output of $AllExpertGF$ as a residual.  Specifically, ablating the $NativeExpertLF$ term would incur a loss on all tasks, with a 0.107\%-0.222\% increase in NE for classification tasks and a 0.158\% increase in MSE for the regression task.

\subsection{Visualization of gating module expert weight distribution}\label{label:visualization}
\begin{figure}[h]
  \centering
  \includegraphics[scale=0.092]{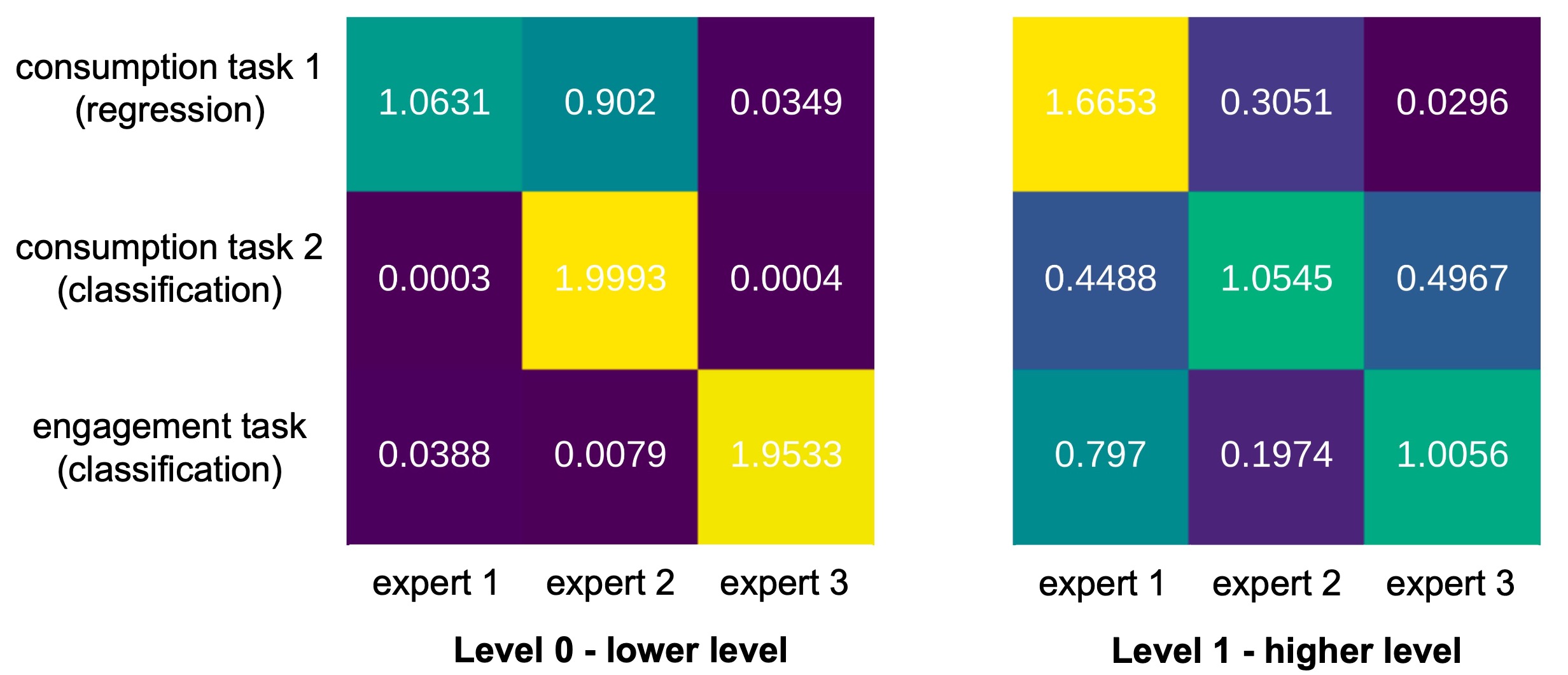}
  \caption{Visualization of the distribution of expert weights learned at each fusion level in a two-level AdaTT-sp. Tasks and experts are arranged in the order of consumption regression task, consumption classification task and engagement task. Note this figure shows the sum of weights from the $NativeExpertLF$ and $AllExpertGF$ modules. As there is only one native expert per task, $NativeExpertLF$ module assigns unit weights to them (mapped to diagonal grids in the figure).}
  \label{fig:gateU}
\end{figure}

In Figure~\ref{fig:gateU}, we visualize the distribution of expert weights after adding weights from both $NativeExpertLF$ and $AllExpertGF$ modules to investigate AdaTT's internal fusion mechanisms. To evaluate the expert utilization, we select three tasks: two consumption tasks and one engagement task. Specifically, we choose one regression task among consumption tasks and two classification tasks with the highest positive event rate among engagement and consumption tasks. We implement two levels of fusion, each with one expert per task and no shared expert. The experts are single-layer MLPs with hidden dimensions of 256 and 128 in the two fusion levels, respectively. After training the model, we apply it to the test dataset, calculate the mean weights across all test examples, and visualize a 3x3 weight matrix for each fusion level. There are some noteworthy observations:

First, at the lower level of fusion (level 0), our model is able to discern relationships between tasks. There is a clear distinction between the consumption and engagement task groups. Additionally, there is an asymmetric sharing pattern among the two consumption tasks: the classification consumption task mostly uses expert 2 and the classification regression task roughly uses expert 1 and 2 equally.

At the higher level of fusion (level 1), where supervision is closer and rich semantic information is captured, our model demonstrates the advantage of soft-parameter sharing through a shared pattern across tasks. While native experts play a significant role for task-specific learning, all experts are utilized flexibly, contributing to shared learning. At this level, the consumption classification task aims to diversify learning by utilizing expert 3, specific to the engagement classification task, as well as expert 1, specific to the consumption regression task. Meanwhile, the engagement task which has fewer positive signals benefits from knowledge transfer from both consumption tasks. In contrast, the consumption regression task primarily relies on its native expert 1 and the expert specific to the other consumption task. Among all experts, expert 1, which has the most diverse learning from a mixture of experts 1 and 2 from level 0, is heavily weighted across all tasks.

In general, we can see clear specialization, where distinct weight distribution patterns are learned at each task, task grouping, and level of fusion.

\begin{table*}
  \caption{Ablation study on the $NativeExpertLF$ module. The performance degradation on every task demonstrates the importance of residual mechanism with separate fusion.}
  \label{tab:ResAblation}
  \begin{tabular}{ccccc}
    \toprule
    Models&Consumption task 1 MSE diff&Consumption task 2 NE diff&Consumption task 3 NE diff&Engagement task 1 NE diff\\
    \midrule
    AdaTT & - & - & - & -\\
    AdaTT (ablated) & +0.158\% & +0.107\% & +0.107\% & +0.222\% \\
   \bottomrule
\end{tabular}
\end{table*}

\begin{table*}
  \caption{AdaTT's performance with different number of experts per task.}
  \label{tab:hyperExp}
  \begin{tabular}{ccccc}
    \toprule
    Number of experts &Consumption task 1 &Consumption task 2 &Consumption task 3 &Engagement task 1\\
    ~ per task & MSE diff&NE diff&NE diff&NE diff\\
    \midrule
    1 & - & - & - & -\\
    2 & -0.171\%&-0.150\% &-0.159\%&-0.075\%\\
    3 & -0.248\%&-0.203\%&-0.239\%&-0.296\%\\
    4 & -0.260\%& -0.230\% & -0.273\% &-0.398\%\\
  \bottomrule
\end{tabular}
\end{table*}

\begin{table*}
  \caption{Results of AdaTT's performance as the fusion level increases. We denote each fusion level's expert hidden dimensions in the first column.}
  \label{tab:FusionLevel}
  \begin{tabular}{ccccc}
    \toprule
    expert &Consumption task 1 &Consumption task 2 &Consumption task 3 &Engagement task 1\\
    hidden dimensions  & MSE diff&NE diff&NE diff&NE diff\\
    \midrule
    256, 128&-&-&- & -\\
    512, 256, 128 & -0.225\%&-0.242\% &-0.284\%&-0.409\%\\
    1024, 512, 256, 128 & -0.503\%&-0.448\%&-0.522\%&-0.619\%\\
    2048, 1024, 512, 256, 128 & -0.664\% &-0.587\% &-0.655\%& -0.766\%\\
  \bottomrule
\end{tabular}
\end{table*}

\subsection{Hyperparameter studies}\label{label:hyperparameters}
We conduct hyperparameter studies to investigate the impact of the number of experts and the number of fusion levels. Both studies use 5 prediction tasks similar to Section~\ref{label:FiveTasks}, with 70 billion examples for training and 10 billion examples for testing. In both studies, we employ AdaTT-sp as the model.

\subsubsection{Effect of number of task-specific experts}
In order to examine the effect of the number of task-specific experts, we keep the number of task-specific experts consistent across all tasks for simplicity and vary it between 1 and 4. These experts are constructed using one-layer MLPs with hidden dimensions of 256 and 128 across two fusion levels. The results of this analysis can be found in Table~\ref{tab:hyperExp}. We can observe that all tasks have improved performance, as the number of experts increases. Nevertheless, the improvements are not consistent: in this study, when the number of experts is increased to 2, the engagement task only exhibits a minor improvement in NE compared to the consumption tasks. However, as the number of experts further increases to 3 and 4, the trend reverses, and the engagement task demonstrates a more noticeable difference in metrics.

\subsubsection{Effect of fusion levels}
We also examine various configurations of fusion levels by utilizing a single expert per task. We incrementally increase the number of fusion levels and use single-layer MLPs for each level. We train four models with hidden dimensions of [256, 128], [512, 256, 128], [1024, 512, 256, 128], and [2048, 1024, 512, 256, 128] for each MLP expert at different fusion levels. For task towers, each model utilizes single-layer MLPs with a hidden dimension of 64. The results are presented in Table~\ref{tab:FusionLevel}. As expected, adding more fusion levels results in greater performance gains. Even when the number of fusion levels is increased to five, considerable improvements are still observed across all tasks.

\section{Conclusion}
In this work, we propose a new  MTL model called Adaptive Task-to-Task Fusion Network (AdaTT). By leveraging its adaptive fusion mechanism, AdaTT effectively models complex task relationships and facilitates the joint learning of task-specific and shared knowledge. Through comprehensive evaluations on both a real-world industrial dataset with diverse task groups, as well as a public dataset, we demonstrate the effectiveness and generalizability of AdaTT. Our results show that AdaTT outperforms state-of-the-art multi-task learning models by a significant margin. We hope to see our work benefits a broader range of applications beyond multitask learning, where different relevant specialized modules can learn synergically.

\bibliographystyle{ACM-Reference-Format}
\balance
\bibliography{ref}


\begin{thebibliography}{36}


\ifx \showCODEN    \undefined \def \showCODEN     #1{\unskip}     \fi
\ifx \showDOI      \undefined \def \showDOI       #1{#1}\fi
\ifx \showISBNx    \undefined \def \showISBNx     #1{\unskip}     \fi
\ifx \showISBNxiii \undefined \def \showISBNxiii  #1{\unskip}     \fi
\ifx \showISSN     \undefined \def \showISSN      #1{\unskip}     \fi
\ifx \showLCCN     \undefined \def \showLCCN      #1{\unskip}     \fi
\ifx \shownote     \undefined \def \shownote      #1{#1}          \fi
\ifx \showarticletitle \undefined \def \showarticletitle #1{#1}   \fi
\ifx \showURL      \undefined \def \showURL       {\relax}        \fi
\providecommand\bibfield[2]{#2}
\providecommand\bibinfo[2]{#2}
\providecommand\natexlab[1]{#1}
\providecommand\showeprint[2][]{arXiv:#2}

\bibitem[Aoki et~al\mbox{.}(2021)]%
        {AokiBIOKDD21}
\bibfield{author}{\bibinfo{person}{Raquel Aoki}, \bibinfo{person}{Frederick
  Tung}, {and} \bibinfo{person}{Gabriel~L. Oliveira}.}
  \bibinfo{year}{2021}\natexlab{}.
\newblock \showarticletitle{{Heterogeneous Multi-task Learning with Expert
  Diversity}}. In \bibinfo{booktitle}{\emph{BIOKDD}}.
\newblock


\bibitem[Caruana(1997)]%
        {caruana1997multitask}
\bibfield{author}{\bibinfo{person}{Rich Caruana}.}
  \bibinfo{year}{1997}\natexlab{}.
\newblock \showarticletitle{Multitask learning}.
\newblock \bibinfo{journal}{\emph{Machine learning}} \bibinfo{volume}{28},
  \bibinfo{number}{1} (\bibinfo{year}{1997}), \bibinfo{pages}{41--75}.
\newblock


\bibitem[Chen et~al\mbox{.}(2018)]%
        {chen2018gradnorm}
\bibfield{author}{\bibinfo{person}{Zhao Chen}, \bibinfo{person}{Vijay
  Badrinarayanan}, \bibinfo{person}{Chen-Yu Lee}, {and} \bibinfo{person}{Andrew
  Rabinovich}.} \bibinfo{year}{2018}\natexlab{}.
\newblock \showarticletitle{Gradnorm: Gradient normalization for adaptive loss
  balancing in deep multitask networks}. In
  \bibinfo{booktitle}{\emph{International conference on machine learning}}.
  PMLR, \bibinfo{pages}{794--803}.
\newblock


\bibitem[Chen et~al\mbox{.}(2020)]%
        {chen2020just}
\bibfield{author}{\bibinfo{person}{Zhao Chen}, \bibinfo{person}{Jiquan Ngiam},
  \bibinfo{person}{Yanping Huang}, \bibinfo{person}{Thang Luong},
  \bibinfo{person}{Henrik Kretzschmar}, \bibinfo{person}{Yuning Chai}, {and}
  \bibinfo{person}{Dragomir Anguelov}.} \bibinfo{year}{2020}\natexlab{}.
\newblock \showarticletitle{Just pick a sign: Optimizing deep multitask models
  with gradient sign dropout}.
\newblock \bibinfo{journal}{\emph{Advances in Neural Information Processing
  Systems}}  \bibinfo{volume}{33} (\bibinfo{year}{2020}),
  \bibinfo{pages}{2039--2050}.
\newblock


\bibitem[Collobert and Weston(2008)]%
        {collobert2008unified}
\bibfield{author}{\bibinfo{person}{Ronan Collobert} {and}
  \bibinfo{person}{Jason Weston}.} \bibinfo{year}{2008}\natexlab{}.
\newblock \showarticletitle{A unified architecture for natural language
  processing: Deep neural networks with multitask learning}. In
  \bibinfo{booktitle}{\emph{Proceedings of the 25th international conference on
  Machine learning}}. \bibinfo{pages}{160--167}.
\newblock


\bibitem[Deng et~al\mbox{.}(2013)]%
        {deng2013new}
\bibfield{author}{\bibinfo{person}{Li Deng}, \bibinfo{person}{Geoffrey Hinton},
  {and} \bibinfo{person}{Brian Kingsbury}.} \bibinfo{year}{2013}\natexlab{}.
\newblock \showarticletitle{New types of deep neural network learning for
  speech recognition and related applications: An overview}. In
  \bibinfo{booktitle}{\emph{2013 IEEE international conference on acoustics,
  speech and signal processing}}. IEEE, \bibinfo{pages}{8599--8603}.
\newblock


\bibitem[Dua and Graff(2017)]%
        {Dua:2019}
\bibfield{author}{\bibinfo{person}{Dheeru Dua} {and} \bibinfo{person}{Casey
  Graff}.} \bibinfo{year}{2017}\natexlab{}.
\newblock \bibinfo{title}{{UCI} Machine Learning Repository}.
\newblock
\newblock
\urldef\tempurl%
\url{http://archive.ics.uci.edu/ml}
\showURL{%
\tempurl}


\bibitem[Elsken et~al\mbox{.}(2017)]%
        {elsken2017simple}
\bibfield{author}{\bibinfo{person}{Thomas Elsken}, \bibinfo{person}{Jan-Hendrik
  Metzen}, {and} \bibinfo{person}{Frank Hutter}.}
  \bibinfo{year}{2017}\natexlab{}.
\newblock \showarticletitle{Simple and efficient architecture search for
  convolutional neural networks}.
\newblock \bibinfo{journal}{\emph{arXiv preprint arXiv:1711.04528}}
  (\bibinfo{year}{2017}).
\newblock


\bibitem[Guo et~al\mbox{.}(2020)]%
        {guo2020learning}
\bibfield{author}{\bibinfo{person}{Pengsheng Guo}, \bibinfo{person}{Chen-Yu
  Lee}, {and} \bibinfo{person}{Daniel Ulbricht}.}
  \bibinfo{year}{2020}\natexlab{}.
\newblock \showarticletitle{Learning to branch for multi-task learning}. In
  \bibinfo{booktitle}{\emph{International Conference on Machine Learning}}.
  PMLR, \bibinfo{pages}{3854--3863}.
\newblock


\bibitem[Hadash et~al\mbox{.}(2018)]%
        {hadash2018rank}
\bibfield{author}{\bibinfo{person}{Guy Hadash}, \bibinfo{person}{Oren~Sar
  Shalom}, {and} \bibinfo{person}{Rita Osadchy}.}
  \bibinfo{year}{2018}\natexlab{}.
\newblock \showarticletitle{Rank and rate: multi-task learning for recommender
  systems}. In \bibinfo{booktitle}{\emph{Proceedings of the 12th ACM Conference
  on Recommender Systems}}. \bibinfo{pages}{451--454}.
\newblock


\bibitem[Hashimoto et~al\mbox{.}(2016)]%
        {hashimoto2016joint}
\bibfield{author}{\bibinfo{person}{Kazuma Hashimoto}, \bibinfo{person}{Caiming
  Xiong}, \bibinfo{person}{Yoshimasa Tsuruoka}, {and} \bibinfo{person}{Richard
  Socher}.} \bibinfo{year}{2016}\natexlab{}.
\newblock \showarticletitle{A joint many-task model: Growing a neural network
  for multiple nlp tasks}.
\newblock \bibinfo{journal}{\emph{arXiv preprint arXiv:1611.01587}}
  (\bibinfo{year}{2016}).
\newblock


\bibitem[He et~al\mbox{.}(2016)]%
        {he2016deep}
\bibfield{author}{\bibinfo{person}{Kaiming He}, \bibinfo{person}{Xiangyu
  Zhang}, \bibinfo{person}{Shaoqing Ren}, {and} \bibinfo{person}{Jian Sun}.}
  \bibinfo{year}{2016}\natexlab{}.
\newblock \showarticletitle{Deep residual learning for image recognition}. In
  \bibinfo{booktitle}{\emph{Proceedings of the IEEE conference on computer
  vision and pattern recognition}}. \bibinfo{pages}{770--778}.
\newblock


\bibitem[He et~al\mbox{.}(2014)]%
        {he2014practical}
\bibfield{author}{\bibinfo{person}{Xinran He}, \bibinfo{person}{Junfeng Pan},
  \bibinfo{person}{Ou Jin}, \bibinfo{person}{Tianbing Xu}, \bibinfo{person}{Bo
  Liu}, \bibinfo{person}{Tao Xu}, \bibinfo{person}{Yanxin Shi},
  \bibinfo{person}{Antoine Atallah}, \bibinfo{person}{Ralf Herbrich},
  \bibinfo{person}{Stuart Bowers}, {et~al\mbox{.}}}
  \bibinfo{year}{2014}\natexlab{}.
\newblock \showarticletitle{Practical lessons from predicting clicks on ads at
  facebook}. In \bibinfo{booktitle}{\emph{Proceedings of the eighth
  international workshop on data mining for online advertising}}.
  \bibinfo{pages}{1--9}.
\newblock


\bibitem[Javaloy and Valera(2021)]%
        {javaloy2021rotograd}
\bibfield{author}{\bibinfo{person}{Adri{\'a}n Javaloy} {and}
  \bibinfo{person}{Isabel Valera}.} \bibinfo{year}{2021}\natexlab{}.
\newblock \showarticletitle{RotoGrad: Gradient Homogenization in Multitask
  Learning}.
\newblock \bibinfo{journal}{\emph{arXiv preprint arXiv:2103.02631}}
  (\bibinfo{year}{2021}).
\newblock


\bibitem[Kendall et~al\mbox{.}(2018)]%
        {kendall2018multi}
\bibfield{author}{\bibinfo{person}{Alex Kendall}, \bibinfo{person}{Yarin Gal},
  {and} \bibinfo{person}{Roberto Cipolla}.} \bibinfo{year}{2018}\natexlab{}.
\newblock \showarticletitle{Multi-task learning using uncertainty to weigh
  losses for scene geometry and semantics}. In
  \bibinfo{booktitle}{\emph{Proceedings of the IEEE conference on computer
  vision and pattern recognition}}. \bibinfo{pages}{7482--7491}.
\newblock


\bibitem[Kokkinos(2017)]%
        {kokkinos2017ubernet}
\bibfield{author}{\bibinfo{person}{Iasonas Kokkinos}.}
  \bibinfo{year}{2017}\natexlab{}.
\newblock \showarticletitle{Ubernet: Training a universal convolutional neural
  network for low-, mid-, and high-level vision using diverse datasets and
  limited memory}. In \bibinfo{booktitle}{\emph{Proceedings of the IEEE
  conference on computer vision and pattern recognition}}.
  \bibinfo{pages}{6129--6138}.
\newblock


\bibitem[Liu et~al\mbox{.}(2018b)]%
        {liu2018progressive}
\bibfield{author}{\bibinfo{person}{Chenxi Liu}, \bibinfo{person}{Barret Zoph},
  \bibinfo{person}{Maxim Neumann}, \bibinfo{person}{Jonathon Shlens},
  \bibinfo{person}{Wei Hua}, \bibinfo{person}{Li-Jia Li}, \bibinfo{person}{Li
  Fei-Fei}, \bibinfo{person}{Alan Yuille}, \bibinfo{person}{Jonathan Huang},
  {and} \bibinfo{person}{Kevin Murphy}.} \bibinfo{year}{2018}\natexlab{b}.
\newblock \showarticletitle{Progressive neural architecture search}. In
  \bibinfo{booktitle}{\emph{Proceedings of the European conference on computer
  vision (ECCV)}}. \bibinfo{pages}{19--34}.
\newblock


\bibitem[Liu et~al\mbox{.}(2018a)]%
        {liu2018darts}
\bibfield{author}{\bibinfo{person}{Hanxiao Liu}, \bibinfo{person}{Karen
  Simonyan}, {and} \bibinfo{person}{Yiming Yang}.}
  \bibinfo{year}{2018}\natexlab{a}.
\newblock \showarticletitle{Darts: Differentiable architecture search}.
\newblock \bibinfo{journal}{\emph{arXiv preprint arXiv:1806.09055}}
  (\bibinfo{year}{2018}).
\newblock


\bibitem[Liu et~al\mbox{.}(2019)]%
        {liu2019end}
\bibfield{author}{\bibinfo{person}{Shikun Liu}, \bibinfo{person}{Edward Johns},
  {and} \bibinfo{person}{Andrew~J Davison}.} \bibinfo{year}{2019}\natexlab{}.
\newblock \showarticletitle{End-to-end multi-task learning with attention}. In
  \bibinfo{booktitle}{\emph{Proceedings of the IEEE/CVF conference on computer
  vision and pattern recognition}}. \bibinfo{pages}{1871--1880}.
\newblock


\bibitem[Long et~al\mbox{.}(2017)]%
        {long2017learning}
\bibfield{author}{\bibinfo{person}{Mingsheng Long}, \bibinfo{person}{Zhangjie
  Cao}, \bibinfo{person}{Jianmin Wang}, {and} \bibinfo{person}{Philip~S Yu}.}
  \bibinfo{year}{2017}\natexlab{}.
\newblock \showarticletitle{Learning multiple tasks with multilinear
  relationship networks}.
\newblock \bibinfo{journal}{\emph{Advances in neural information processing
  systems}}  \bibinfo{volume}{30} (\bibinfo{year}{2017}).
\newblock


\bibitem[Ma et~al\mbox{.}(2019)]%
        {ma2019snr}
\bibfield{author}{\bibinfo{person}{Jiaqi Ma}, \bibinfo{person}{Zhe Zhao},
  \bibinfo{person}{Jilin Chen}, \bibinfo{person}{Ang Li},
  \bibinfo{person}{Lichan Hong}, {and} \bibinfo{person}{Ed~H Chi}.}
  \bibinfo{year}{2019}\natexlab{}.
\newblock \showarticletitle{Snr: Sub-network routing for flexible parameter
  sharing in multi-task learning}. In \bibinfo{booktitle}{\emph{Proceedings of
  the AAAI Conference on Artificial Intelligence}}, Vol.~\bibinfo{volume}{33}.
  \bibinfo{pages}{216--223}.
\newblock


\bibitem[Ma et~al\mbox{.}(2018)]%
        {ma2018modeling}
\bibfield{author}{\bibinfo{person}{Jiaqi Ma}, \bibinfo{person}{Zhe Zhao},
  \bibinfo{person}{Xinyang Yi}, \bibinfo{person}{Jilin Chen},
  \bibinfo{person}{Lichan Hong}, {and} \bibinfo{person}{Ed~H Chi}.}
  \bibinfo{year}{2018}\natexlab{}.
\newblock \showarticletitle{Modeling task relationships in multi-task learning
  with multi-gate mixture-of-experts}. In \bibinfo{booktitle}{\emph{Proceedings
  of the 24th ACM SIGKDD international conference on knowledge discovery \&
  data mining}}. \bibinfo{pages}{1930--1939}.
\newblock


\bibitem[Meyerson and Miikkulainen(2017)]%
        {meyerson2017beyond}
\bibfield{author}{\bibinfo{person}{Elliot Meyerson} {and}
  \bibinfo{person}{Risto Miikkulainen}.} \bibinfo{year}{2017}\natexlab{}.
\newblock \showarticletitle{Beyond shared hierarchies: Deep multitask learning
  through soft layer ordering}.
\newblock \bibinfo{journal}{\emph{arXiv preprint arXiv:1711.00108}}
  (\bibinfo{year}{2017}).
\newblock


\bibitem[Misra et~al\mbox{.}(2016)]%
        {misra2016cross}
\bibfield{author}{\bibinfo{person}{Ishan Misra}, \bibinfo{person}{Abhinav
  Shrivastava}, \bibinfo{person}{Abhinav Gupta}, {and} \bibinfo{person}{Martial
  Hebert}.} \bibinfo{year}{2016}\natexlab{}.
\newblock \showarticletitle{Cross-stitch networks for multi-task learning}. In
  \bibinfo{booktitle}{\emph{Proceedings of the IEEE conference on computer
  vision and pattern recognition}}. \bibinfo{pages}{3994--4003}.
\newblock


\bibitem[Real et~al\mbox{.}(2019)]%
        {real2019regularized}
\bibfield{author}{\bibinfo{person}{Esteban Real}, \bibinfo{person}{Alok
  Aggarwal}, \bibinfo{person}{Yanping Huang}, {and} \bibinfo{person}{Quoc~V
  Le}.} \bibinfo{year}{2019}\natexlab{}.
\newblock \showarticletitle{Regularized evolution for image classifier
  architecture search}. In \bibinfo{booktitle}{\emph{Proceedings of the aaai
  conference on artificial intelligence}}, Vol.~\bibinfo{volume}{33}.
  \bibinfo{pages}{4780--4789}.
\newblock


\bibitem[Ruder et~al\mbox{.}(2019)]%
        {ruder2019latent}
\bibfield{author}{\bibinfo{person}{Sebastian Ruder}, \bibinfo{person}{Joachim
  Bingel}, \bibinfo{person}{Isabelle Augenstein}, {and} \bibinfo{person}{Anders
  S{\o}gaard}.} \bibinfo{year}{2019}\natexlab{}.
\newblock \showarticletitle{Latent multi-task architecture learning}. In
  \bibinfo{booktitle}{\emph{Proceedings of the AAAI Conference on Artificial
  Intelligence}}, Vol.~\bibinfo{volume}{33}. \bibinfo{pages}{4822--4829}.
\newblock


\bibitem[Sener and Koltun(2018)]%
        {sener2018multi}
\bibfield{author}{\bibinfo{person}{Ozan Sener} {and} \bibinfo{person}{Vladlen
  Koltun}.} \bibinfo{year}{2018}\natexlab{}.
\newblock \showarticletitle{Multi-task learning as multi-objective
  optimization}.
\newblock \bibinfo{journal}{\emph{Advances in neural information processing
  systems}}  \bibinfo{volume}{31} (\bibinfo{year}{2018}).
\newblock


\bibitem[Sun et~al\mbox{.}(2020)]%
        {sun2020adashare}
\bibfield{author}{\bibinfo{person}{Ximeng Sun}, \bibinfo{person}{Rameswar
  Panda}, \bibinfo{person}{Rogerio Feris}, {and} \bibinfo{person}{Kate
  Saenko}.} \bibinfo{year}{2020}\natexlab{}.
\newblock \showarticletitle{Adashare: Learning what to share for efficient deep
  multi-task learning}.
\newblock \bibinfo{journal}{\emph{Advances in Neural Information Processing
  Systems}}  \bibinfo{volume}{33} (\bibinfo{year}{2020}),
  \bibinfo{pages}{8728--8740}.
\newblock


\bibitem[Tang et~al\mbox{.}(2020)]%
        {tang2020progressive}
\bibfield{author}{\bibinfo{person}{Hongyan Tang}, \bibinfo{person}{Junning
  Liu}, \bibinfo{person}{Ming Zhao}, {and} \bibinfo{person}{Xudong Gong}.}
  \bibinfo{year}{2020}\natexlab{}.
\newblock \showarticletitle{Progressive layered extraction (ple): A novel
  multi-task learning (mtl) model for personalized recommendations}. In
  \bibinfo{booktitle}{\emph{Fourteenth ACM Conference on Recommender Systems}}.
  \bibinfo{pages}{269--278}.
\newblock


\bibitem[Vandenhende et~al\mbox{.}(2019)]%
        {vandenhende2019branched}
\bibfield{author}{\bibinfo{person}{Simon Vandenhende},
  \bibinfo{person}{Stamatios Georgoulis}, \bibinfo{person}{Bert De~Brabandere},
  {and} \bibinfo{person}{Luc Van~Gool}.} \bibinfo{year}{2019}\natexlab{}.
\newblock \showarticletitle{Branched multi-task networks: deciding what layers
  to share}.
\newblock \bibinfo{journal}{\emph{arXiv preprint arXiv:1904.02920}}
  (\bibinfo{year}{2019}).
\newblock


\bibitem[Wang et~al\mbox{.}(2022)]%
        {wang2022can}
\bibfield{author}{\bibinfo{person}{Yuyan Wang}, \bibinfo{person}{Zhe Zhao},
  \bibinfo{person}{Bo Dai}, \bibinfo{person}{Christopher Fifty},
  \bibinfo{person}{Dong Lin}, \bibinfo{person}{Lichan Hong},
  \bibinfo{person}{Li Wei}, {and} \bibinfo{person}{Ed~H Chi}.}
  \bibinfo{year}{2022}\natexlab{}.
\newblock \showarticletitle{Can Small Heads Help? Understanding and Improving
  Multi-Task Generalization}. In \bibinfo{booktitle}{\emph{Proceedings of the
  ACM Web Conference 2022}}. \bibinfo{pages}{3009--3019}.
\newblock


\bibitem[Williams et~al\mbox{.}(2008)]%
        {williams2008multi}
\bibfield{author}{\bibinfo{person}{Christopher Williams},
  \bibinfo{person}{Stefan Klanke}, \bibinfo{person}{Sethu Vijayakumar}, {and}
  \bibinfo{person}{Kian Chai}.} \bibinfo{year}{2008}\natexlab{}.
\newblock \showarticletitle{Multi-task gaussian process learning of robot
  inverse dynamics}.
\newblock \bibinfo{journal}{\emph{Advances in neural information processing
  systems}}  \bibinfo{volume}{21} (\bibinfo{year}{2008}).
\newblock


\bibitem[Yu et~al\mbox{.}(2020)]%
        {yu2020gradient}
\bibfield{author}{\bibinfo{person}{Tianhe Yu}, \bibinfo{person}{Saurabh Kumar},
  \bibinfo{person}{Abhishek Gupta}, \bibinfo{person}{Sergey Levine},
  \bibinfo{person}{Karol Hausman}, {and} \bibinfo{person}{Chelsea Finn}.}
  \bibinfo{year}{2020}\natexlab{}.
\newblock \showarticletitle{Gradient surgery for multi-task learning}.
\newblock \bibinfo{journal}{\emph{Advances in Neural Information Processing
  Systems}}  \bibinfo{volume}{33} (\bibinfo{year}{2020}),
  \bibinfo{pages}{5824--5836}.
\newblock


\bibitem[Zhang et~al\mbox{.}(2014)]%
        {zhang2014facial}
\bibfield{author}{\bibinfo{person}{Zhanpeng Zhang}, \bibinfo{person}{Ping Luo},
  \bibinfo{person}{Chen~Change Loy}, {and} \bibinfo{person}{Xiaoou Tang}.}
  \bibinfo{year}{2014}\natexlab{}.
\newblock \showarticletitle{Facial landmark detection by deep multi-task
  learning}. In \bibinfo{booktitle}{\emph{European conference on computer
  vision}}. Springer, \bibinfo{pages}{94--108}.
\newblock


\bibitem[Zhao et~al\mbox{.}(2019)]%
        {zhao2019recommending}
\bibfield{author}{\bibinfo{person}{Zhe Zhao}, \bibinfo{person}{Lichan Hong},
  \bibinfo{person}{Li Wei}, \bibinfo{person}{Jilin Chen},
  \bibinfo{person}{Aniruddh Nath}, \bibinfo{person}{Shawn Andrews},
  \bibinfo{person}{Aditee Kumthekar}, \bibinfo{person}{Maheswaran
  Sathiamoorthy}, \bibinfo{person}{Xinyang Yi}, {and} \bibinfo{person}{Ed
  Chi}.} \bibinfo{year}{2019}\natexlab{}.
\newblock \showarticletitle{Recommending what video to watch next: a multitask
  ranking system}. In \bibinfo{booktitle}{\emph{Proceedings of the 13th ACM
  Conference on Recommender Systems}}. \bibinfo{pages}{43--51}.
\newblock


\bibitem[Zoph and Le(2016)]%
        {zoph2016neural}
\bibfield{author}{\bibinfo{person}{Barret Zoph} {and} \bibinfo{person}{Quoc~V
  Le}.} \bibinfo{year}{2016}\natexlab{}.
\newblock \showarticletitle{Neural architecture search with reinforcement
  learning}.
\newblock \bibinfo{journal}{\emph{arXiv preprint arXiv:1611.01578}}
  (\bibinfo{year}{2016}).
\newblock


\end{thebibliography}

\end{document}